\documentclass[12p]{article}
\usepackage{amsfonts}
\usepackage{amsmath}

\title{Credit Value Adjustment for Counterparties\\ with Illiquid CDS}
\author{Ola Hammarlid\\Marta Leniec}

\begin{document}

\maketitle
\abstract{Credit Value Adjustment (CVA) is the difference between the value of the default-free and credit-risky derivative portfolio, which can be regarded as the cost of the credit hedge. Default probabilities are therefore needed, as input parameters to the valuation. When liquid CDS are available, then implied probabilities of default can be derived and used. However,  in small markets, like the Nordic region of Europe, there are practically no CDS to use. We study the following problem: given that no liquid contracts written on the default event are available, choose a model for the default time and estimate the model parameters. We use the minimum variance hedge to show that we should use the real-world probabilities, first in a discrete time setting and later in the  continuous time setting. We also argue that this approach should fulfil the requirements of IFRS 13, which means it could be used in accounting as well. We also present a method that can be used to estimate the real-world probabilities of default, making maximal use of market information (IFRS requirement). 

\section{Introduction}
Credit Value Adjustment (CVA) is defined as the difference in value of a portfolio, without credit risk and when exposed to default risk, see for example (3.1) in Greeen \cite{green} or (12.1) in Gregory \cite{gregory}. This difference can be regarded as a derivative, and as such its value is equal to the cost of the hedge. 

The Probability of Default (PD) and Loss Given Default (LGD) are two of the most important parameters that impact CVA. If derivatives of the default event exist, like Credit Default Swaps (CDS), then implied probability of default can be derived and used in pricing. Our aim is to answer the following question: How should we model and price the credit risk of derivatives (CVA) for counterparties for which there does not exist any derivatives that can be used for hedging default risk.

We argue that in that case it is not the extended implied methodology (e.g. Nomura model in \cite{c5}) that should be used. The basic reasoning is based on minimum variance hedging, which  leads to the use of real-world probabilities.  
To illustrate our line of argument, when a derivatives on the default is lacking, we start with a discrete time example. 

\subsection{An introductory example}
\label{discrete}
In this introductory example of variance minimization, the time of default  $\tau$ can only occur in discrete time $\mathcal{T}=\{t_1, t_2, ...\}$. Let $p_t$ denote the probability that the default is equal $t$, for any $t\in\mathcal{T}$, i.e. \[p_t=\mathbb{P}(\tau = t),\quad\text{for all $t\in\mathcal{T}$}.\]
Let $V_t$ denote the value at time $t$ of a portfolio of derivatives with maturity times less or equal than $T>0$. The positive part of the portfolio is denoted by
\[V_t^+=\max\{V_t,0\},\]
and is called the positive exposure at time $t$. 
The positive exposure at default is therefore given by
\begin{equation}\label{PE_dis}
PE_{\tau}=V_{\tau}^+=\sum_{t\in\mathcal{T},t<T}V_t^+\mathbb{I}_{\tau=t}
.
\end{equation}
CVA is thus the expected value of $V_\tau^+ LGD$,  which is the cost of hedge of this entity.  
Moreover, for later mathematical convenience we denote the following vector
\[\mathcal{V}=[V_{t_1}^+,V_{t_2}^+,...,V_{T}^+].\]
Further, we assume independence between $\tau$ and $V_t$ for each $t$, which is a common assumption for CVA. It is however not necessary, but give better clarity of the concept and hopefully better intuition.  

We assume that the default-free market is arbitrage-free and complete and consequently every $V_t^+$ can be perfectly replicated. Pricing is determined by the expectation of the pay-off under the unique equivalent martingale measure $\mathbb{Q}^*$. Moreover, in the case of no liquid CDS, the defaultable market is arbitrage-free and incomplete, since the pay-offs of defaultable contracts cannot be perfectly replicated. Hence, perfect hedge is not possible and one needs to use another hedging strategy.

We use the minimum variance hedging to find the hedge and its value. Specifically,  let $\mathcal{H}$ denote the set of possible positions in hedging strategies. Every $h\in\mathcal{H}$ can be written as
\[h=[h_{t_1},h_{t_2},...,h_T],\]
where each $h_{t_i}$ denotes the number of contracts with pay off $V_{t_i}^+.$ Then, the minimum variance hedging portfolio problem can be written as
\begin{equation}\label{opt}h^*=argmin_{h\in\mathcal{H}}Var\left[PE_\tau-h\cdot\mathcal{V}\right],\end{equation}
where $Var$ denotes the variance under the real-world probability measure $\mathbb{P}$. We use the real-world probability measure $\mathbb{P}$ because risk assessments, such as variance, are in the real-word probabilities, since gains and losses are real-world entities.

By Proposition 3.2 in Hult et al. \cite{c13} we get that the unique solution of the minimization problem (\ref{opt})
is given by
\[h^*=\Sigma_{\mathcal{V}}^{-1}\Sigma_{PE,\mathcal{V}},\]
where
\begin{equation}
\Sigma_{\mathcal{V}}^{-1} = 
\begin{pmatrix}Cov(V_{t_1}^+,V_{t_1}^+) & ... & Cov(V_{t_1}^+,V_{T}^+)\\ ... & ... & ...\\ Cov(V_{T}^+,V_{t_1}^+) & ... &Cov(V_{T}^+,V_{T}^+) \end{pmatrix}^{-1}
\end{equation}
and
\begin{equation}
\Sigma_{PE,\mathcal{V}} = 
\begin{pmatrix}Cov(PE,V_{t_1}^+) \\...\\ Cov(PE,V_{T}^+) \end{pmatrix}.
\end{equation}
By (\ref{PE_dis}) we have that
\begin{equation}
\begin{split}
Cov\left(PE,V_{t_i}^+\right)&=Cov\left(\sum_{t\in\mathcal{T},t<T}V_t^+\mathbb{I}_{\tau=t},V_{t_i}^+\right)\\
&=\sum_{t=t_1}^T \mathbb{E}^{\mathbb{P}}\left[V_{t}^+V_{t_i}^+\right]\cdot p_t-\sum_{t=t_1}^T\mathbb{E}^{\mathbb{P}}\left[V_{t}^+\right]\mathbb{E}^{\mathbb{P}}\left[V_{t_i}^+\right]\cdot p_t\\
&=\sum_{t=t_1}^T \left(\mathbb{E}^{\mathbb{P}}\left[V_{t}^+V_{t_i}^+\right]-\mathbb{E}^{\mathbb{P}}\left[V_{t}^+\right]\mathbb{E}^{\mathbb{P}}\left[V_{t_i}^+\right]\right)\cdot p_t\\
&=\sum_{t=t_1}^T Cov(V_{t}^+,V_{t_i}^+)\cdot p_t,
\end{split}
\end{equation}
where the second equality is by the assumption that the default risk and the market risk are independent.

Hence, we have\[\Sigma_{PE,\mathcal{V}}=\Sigma_{\mathcal{V}}\cdot\mathcal{P},\]
where
\[\mathcal{P}=(p_{t_1},p_{t_2},...,p_T)^T.\]
Finally, we get that
\begin{equation}
\begin{split}
h^*=\Sigma_{\mathcal{V}}^{-1}\Sigma_{PE,\mathcal{V}}=\Sigma_{\mathcal{V}}^{-1}\cdot\Sigma_{\mathcal{V}}\cdot\mathcal{P}=\mathcal{P},
\end{split}
\end{equation}
which imply that the hedging strategies are given by $\mathcal{P}.$ We note that we would get the same hedging strategy if we took the variance under the minimal martingale measure (see Section \ref{imperfect}).

Consequently, the CVA of the contract is equal to the cost of the hedging portfolio which is equal to
\[CVA=\sum_{t=t_1}^T \mathbb{E}^{\mathbb{Q}^*}\left[e^{-rt}V_{t}^*\right]\cdot p_t,\]
because the value of the contract with payoff $V_{t}^*$ is given by $\mathbb{E}^{\mathbb{Q}^*}\left[e^{-rt}V_{t}^*\right]$ in the default-free market, where $r$ is a deterministic and constant interest rate.

\subsection{Minimum variance hedging}

There are several models of default, where the most popular approaches are either in the category the structural approach or the reduced-form approach (see for example \cite{c20}). We consider the latter and  assume that the default time is an exponentially distributed random variable with a piecewise-constant intensity. 

A commonly used method to estmimate the implied probability of default, when CDS does not exist, is proposed by Nomura in \cite{c5}. 
In this model CDS spreads of liquid names are used to construct proxy CDS spreads for illiquid names,  by a mapping process (for a presentation of the model see Appendix \ref{nomura}). However, this method is based on the assumption that a CDS spread exists and results in implied default probabilities. In our opinion we should build the model from a hedging assumption, using only the available contracts and not hypothetical ones. 

As it was pointed out in Green, the number of credit default swaps in the US market is much larger than in other regions and therefore one could question how appropriate the Nomura method is.  Green mentions that the XVA trading desks should understand that if the proxy CDS is used for hedging, it will not be effective at the actual default time and hence it does not hedge the default risk. Moreover, Green claims in \cite{green} that "risk warehousing is inevitable and this leads directly to incomplete markets and the physical measure". This article supports Green's claim. 

Furthermore, in markets like the Northern Europe, the number of CDS is negligible and therefore it is sub-optimal to model the default probabilities of the majority of counterparties based on such a small sample.  Especially, since CVA is a portfolio effect that is different from a stand alone derivative. 

Since perfect hedging is not possible in incomplete markets, the common approach in the literature is to determine a hedging policy according to some criterion. Starting with the Markowitz optimal portfolio selection (see \cite{c14}), the variance-minimizing criterion has widely been employed in the literature in various economic contexts. The main references for the general case of hedging in the incomplete markets are  Hull \cite{c17}, McDonald \cite{c16} and Stulz \cite{c15}.

Moreover, F\"ollmer and Sondermannn  \cite{f-s} and then F\"ollmer and Schweizer in \cite{c10} presented the connection of variance hedging and the minimal martingale measure, where the minimal martingale measure preserves the structure of the real-world measure as far as possible, under the constraint that the discounted underlying stock price is a martingale. It was discussed in \cite{f-sch} that the decomposition of any contingent claim under the minimal martingale measure provides the so-called F\"ollmer-Schweizer (see for example \cite{c10}) decomposition of the contingent claims under the real-world measure, and this in turn immediately gives the variance-minimizing hedging strategy for the claim. In practise it means that if one aims to find a hedging strategy that minimizes the variance of the hedging error, then one should use the minimal martingale measure.

Jeanblanc and Rutkowski \cite{c20} give an overview on default modelling, which is consistent with our approach, and shows that when no defaultable hedging claims exist (i.e. no liquid CDS), the market is incomplete and replication of defaultable claims is not possible. They also suggest the minimum variance hedging of credit derivatives(see  Bielecki, Jeanblanc and Rutkowski \cite{c21} for more details). They also verify that in a complete market, their mean-variance price is equal to the unique arbitrage-free price.

In incomplete markets, there are infinitely many martingale measures which are consistent with the no-arbitrage condition. El Karoui et al  \cite{c7} showed that in case of default risk and incomplete markets, the variance minimization leads to the minimal martingale measure that removes the drift of the underlying stock but leaves the probability of the default unchanged, i.e. the default probability is under the real-world measure. They study the minimal martingale measure approach (mentioned above) as well as the minimal entropy martingale measure

Hence, following this approach, we employ the variance minimization hedging strategy for the problem of pricing and hedging CVA. The interpretation of our result is that an investor who wants to hedge a derivative in the presence of default risk in the incomplete  market, prices the contracts as she was risk-neutral with respect to the default risk. To our knowledge, this line of argument, has so far not been applied before to CVA to support the use of real-world probabilities of default. 

\subsection{The connection to IFRS 13}
This paper studies an applied problem that is present for various banks and financial institutions around the world that fall under the International Financial Reporting Standards (IFRS) regulations. We argue that the requirements presented in IFRS 13 are satisfied by the proposed framework.

Firstly, IFRS 13, \S 2 defines fair value as: The "price at which an orderly transaction to sell the asset or to transfer the liability would take place between market participants at the measurement date under current market conditions (i.e. an exit price at the measurement date from the perspective of a market participant that holds the asset or owes the liability)." The model does not make any assumptions but uses the current state of the world and an orderly trading. 

Secondly, when pricing Credit Valuation Adjustment (CVA), IFRS 13, \S 22 states that:
"An entity shall measure the fair value of an asset or a liability using the assumptions that market participants would use when pricing the asset or liability, assuming that market participants act in their economic best interest." The cost of the hedge is equal to the price of a derivative, hence a price which all participants in the market could agree upon. 

Lastly, when estimating the input parameters that the model consumes IFRS 13, \S 67 states:
Valuation "techniques used to measure fair value shall maximise the use of relevant observable inputs and minimise the use of unobservable inputs." This apply to the second step of the modelling, the estimation of the real-world probabilities of default. 
We present how to make maximum use of market observable information to estimate probability of default.

\section{Pricing and hedging}
\label{mm}
The minimal martingale measure and the minimal entropy martingale measure lead to a particular choice of a  measure from the set of equivalent martingale measures (see El Karoui et al. \cite{c7}). We connect this concepts to the variance-minimizing hedging strategy and consequently CVA.

Let $(\Omega,\mathcal{F},\mathbb{F}=(\mathcal{F}_t)_{t\geq 0},\mathbb{P})$ be a filtered probability space, where $T>0$ is a finite time horizon and $\mathbb{F}$ is the filtration (satisfying the usual conditions) generated by, for example a geometric Brownian motion $W_t$, that is,
\begin{equation}\label{s}
dS_t=\mu(t, S_t) dt+\sigma(t, S_t) dW_t,\quad S_0=s_0\end{equation}
where $\mu$ is the drift and $\sigma>0$ is the volatility and $\mathcal{F}_T\subset\mathcal{F}$. Moreover, let a default time $\tau$ be an exponential random variable with intensity $\lambda^{\mathbb{P}}>0$ defined on $(\Omega,\mathcal{F})$ and denote
by $\mathbb{G}=(\mathcal{G})_{t\in[0,T]}$ a filtration (satisfying the usual conditions) such that
\[\mathcal{G}_t=\mathcal{F}_t\vee\sigma(\tau\wedge t)\quad\text{for all $t\in[0,T]$}.\]

The stock price uncertainty is market risk and the uncertainty coming from the default is default risk. 
We still assume independence between $\tau$ and $W_t$ for all $t\geq 0. $ 

The filtration $\mathbb{G}=(\mathcal{G}_t)_{t\geq 0}$ represents the information  $S$ and the information about the default time $\tau,$ i.e. at any time $t\geq 0$ we have observed the price up to time $t\geq 0$ and we know whether $\tau$ has already occurred or not. This is a standard way of modelling information flow in the area of financial mathematics and more details can be found for example in \cite{c1} and \cite{c2}. Moreover, the practitioners use this way of modelling the information level in case of default risk. The standard reference is \cite{green} which is the handbook for CVA calculations.

\subsection{Default-free market}
\label{def-free}
Let us begin with introducing a default-free market consisting of a price $S$ defined by (\ref{s}) and a bank account $B=(B_t)_{t\in[0,T]},$ where
\[dB_t = rB_tdt,\quad B_0=1\]
and $r$ is a constant interest-rate. Since the coefficients of the geometric Brownian motion $S$ are constant and $\sigma>0,$ the Assumption 1 and Assumption 2 of Blanchet-Scalliet et al. \cite{c7} are satisfied and the default-free market is complete and arbitrage-free. For some more details see for example Karatzas \cite{c9}.

The information flow in this case is the filtration $\mathbb{F}$ generated by the  price $S$ and the unique equivalent martingale measure $\mathbb{Q}^*$ on $\mathbb{F}$ is given by
\[d\mathbb{Q}^*|_{\mathcal{F}_t} = Z^*_t d\mathbb{P}|_{\mathcal{F}_t}\quad \text{for all $t\in[0,T]$},\]
where $Z^*=(Z^*_t)_{t\in[0,T]}$ is the Radon-Nikodym derivative of $\mathbb{Q}^*$ with respect to $\mathbb{P}$ given by
\begin{equation}\label{z}
Z^*_t=\exp\left \{-\frac{\theta^2}{2}t+\theta W_t \right \},\quad Z_0^*=1\end{equation}
and $\theta=-(\mu-r)\sigma^{-1}.$
We see that the only source of randomness in this market is the market risk coming from the Brownian motion $W,$ i.e. from the fluctuations of the price $S.$

Let us denote by $\mathcal{M}(\mathbb{F})$ the set of equivalent martingale measures on $\mathbb{F}.$ Then we have that
\[\mathcal{M}(\mathbb{F})=\{\mathbb{Q}^*\}.\]
Consequently, any $\mathcal{F}_T$-measurable contingent claim $X_T$ has a unique price given by
\[\mathbb{E}^{\mathbb{Q}^*}\left[e^{-rT}X_T\right]=\mathbb{E}^{\mathbb{P}}\left[e^{-rT}X_TZ_T^*\right].\]

\subsection{Defaultable market}
Now we extend the default-free market defined in Subsection \ref{def-free} by introducing the random default time $\tau$ which is exponentially distributed with an intensity $\lambda^{\mathbb{P}}>0.$ As discussed above, the information level in this case is given by the filtration $\mathbb{G}$ and thus the set of equivalent measures making the discounted price process a martingale has to be defined on the filtration $\mathbb{G}$ (see for example \cite{c1}).

Since we assume that the default time and the  price are independent, the Assumption 3 of \cite{c7} is satisfied. Moreover, the exponential distribution assumption of $\tau$ makes the Assumption 4 and Assumption 5 of \cite{c7} satisfied.

Let $\mathcal{M}(\mathbb{G})$ denote the set of equivalent martingale measures for the filtration $\mathbb{G}.$ Since we assumed that $\tau$ is independent of the Brownian motion $W,$ the Jacod's hypothesis is satisfied (see for example \cite{c1}  for the Jacod's hypothesis) and consequently there exists at least one equivalent martingale measure in $\mathcal{M}(\mathbb{G}).$

As shown in \cite{c7}, \cite{c8}, \cite{c3}  and other articles in the area of credit risk, if $\mathbb{Q}^H\in\mathcal{M}(\mathbb{G}),$ then the Radon-Nikodym derivative $L^H=(L^H_t)_{t\in[0,T]}$ for the change of probability measure from $\mathbb{Q}^H$ to $\mathbb{P}$ on $\mathbb{G}$ is given by
\begin{equation}\label{l_gen}L^H_t=L_0(t)\cdot L^H_1(t),\end{equation}
where
\[L_0(t)=\exp\left\{-\frac{\theta^2}{2}t+\theta W_t \right \}\]
and
\[L^H_1(t)=\exp\left\{H_{\tau}\mathbb{I}_{\tau\leq t}-\lambda^{\mathbb{P}}\int_0^{t\wedge t}\left(e^{H_s}-1\right)ds\right\},\]
where $H=(H_t)_{t\in[0,T]}$ is a $\mathbb{G}$-adapted process satisfying some technical conditions (see for example \cite{c3}).

Moreover, it was shown for example in \cite{c7} that the intensity $\lambda_t^{\mathbb{Q}^H}$ of the default time $\tau$ under measure $\mathbb{Q}^H$ satisfies
\[\lambda_t^{\mathbb{Q}^H}=e^{H_t}\cdot\lambda^{\mathbb{P}}\quad\text{for any $t\in[0,T].$}\]

We see that $L_0(t)$ is equal to the Radon-Nikodym derivative for the unique equivalent change of measure from $\mathbb{Q}^*$ to $\mathbb{P}$ on $\mathbb{F}$ and since $\tau$ is independent of $W_t$ then $L_0(t)$ is independent of $\tau.$

Since $H$ is a $\mathbb{G}$-adapted process, then in general $L_1^H(t)$ is not independent of $W_t.$ Also, in general $L_1^H(t)$ is not independent of $\tau.$
\subsubsection{Incompleteness of the defaultable market}
\label{incompet}
It was shown for example in \cite{c8} and \cite{c3} that the set $\mathcal{M}(\mathbb{G})$ has infinitely many elements, which means that introducing the default risk to the default-free market brings some form of incompleteness. Specifically, in \cite{c3} the authors show that the measure $\mathbb{Q}^0$ (i.e. a measure from $\mathcal{M}(\mathbb{G})$ such that $H_t=0$ for any $t\in[0,T]$) belongs to $\mathcal{M}(\mathbb{G})$ and the authors in \cite{c8} show that there exists a measure $\mathbb{Q}^H$ in $\mathcal{M}(\mathbb{G})$ for which $H_t\neq 0$ and hence due to the fact that any convex combination of measures from $\mathcal{M}(\mathbb{G})$ also belongs to $\mathcal{M}(\mathbb{G})$ we get that $\mathcal{M}(\mathbb{G})$  has infinitely many elements. 
As a result, one faces the problem of narrowing the set of equivalent martingale measures. 

One method for narrowing down the set $\mathcal{M}(\mathbb{G})$ would be completing the defaultable market by introducing so-called generalized risk-free assets. A generalised risk-free asset would be for example an asset paying $1$ at the default time $\tau,$ i.e. a CDS can be one of them. Then, if such contracts were dynamically traded, then one would be able to extract the risk-neutral intensity $\lambda^{\mathbb{Q}^H}$ for a particular counterparty from the prices of these contracts and use it to calculate CVA. This intensity may be regarded as an implied risk-neutral intensity. The authors of \cite{c7} argue that the presence of dynamically traded generalized risk-free assets implies a unique specification of the equivalent martingale measure in $\mathcal{M}(\mathbb{G})$ and hence a complete market. Consequently, if it possible to extract $\lambda^{\mathbb{Q}^H}$ from the prices of the dynamically traded CDS, then $\lambda^{\mathbb{Q}^H}$ should be used for pricing purposes. Hence, if our aim is to calculate CVA of a counterparty with liquid CDS contracts, then we should extract the risk-neutral intensity from the CDS prices (for example by the bootstrap technique) and use it in the CVA calculations. However, if we consider a counterparty without liquid CDS or without any CDS at all, then we should use other techniques for choosing the equivalent martingale measure from the set $\mathcal{M}(\mathbb{G}).$ We present some of these methods in the following subsections.

\subsubsection{Imperfect hedging and the minimal martingale measure}
\label{imperfect}
In this subsection we consider the case of a counterparty without liquid CDS, and as a result, we deal with the problem of narrowing the set of equivalent martingale measures $\mathcal{M}(\mathbb{G})$. 

The connection between the minimal martingale measure and the variance-minimization hedging between the payoff $h(S_{\tau})$ and the terminal wealth generated from a self-financing strategy, was introduced by F\"ollmer and Sondermann in \cite{f-s}. In economical terms; an approximation of the contingent claim in terms of a self-financing strategy with the replication error ("the tracking error") as small as possible.

As in El Karoui et al. \cite{c7}, the imperfect hedging is connected with a minimal martingale measure, which is defined by the following two conditions: an equivalent martingale measure $\mathbb{Q}^H\in\mathcal{M}(\mathbb{G})$ is called minimal martingale measure if $\mathbb{Q}_H=\mathbb{P}$ on $\mathcal{G}_0$ and if every $(\mathbb{P},\mathbb{G})$-square martingale orthogonal to $W$ under $\mathbb{P}$ is a $(\mathbb{Q}^H,\mathbb{G})$-martingale. By Proposition 6 in \cite{c7} we have that the minimal martingale measure is equal to $\mathbb{Q}^0$, i.e. the equivalent martingale measure defined by $H_t=0$ for every $t\in[0,T].$ This corresponds to a zero risk premium associated with the default risk. In other words we get that the pricing measure can be chosen to be the minimal martingale measure $\mathbb{Q}^0$ defined by the Radon-Nikodym derivative $L^0=(L_t^0)_{t\in[0,T]},$ where
\begin{equation}
\label{l}
L_t^0=\exp\left \{-\frac{\theta^2}{2}t+\theta W_t \right \},\quad\text{for any $t\in[0,T].$}
\end{equation}

\section{CVA calculation}
CVA is given by the following formula (see for example $(3.11)$ in Green \cite{green})
\begin{equation}\label{cva}
CVA=\mathbb{E}^{\mathbb{Q}^H}\left[e^{-r\tau}(1-R)V_{\tau}^+\right],\end{equation}
where $\mathbb{Q}^H\in\mathcal{M}(\mathbb{G}),$ the constant $R$ is the recovery rate ($LGD =1-R $) and $V^+$ is the positive exposure.
It is also assumed that $V=(V_t)_{t\in[0,T]}$ is an $\mathbb{F}$-adapted process, i.e. the only uncertainty in $V$ is the randomness coming from the  price $S$ given by (\ref{s}).

Since there do not exist any dynamically traded hedging instruments for the counterparty, we decide on narrowing the set $\mathcal{M}(\mathbb{G})$ by the well-studied method summarized in Subsection \ref{imperfect}. As a result, we use the equivalent martingale measure $\mathbb{Q}^0$ defined by (\ref{l}) and we get that
\[
\begin{split}
CVA=\mathbb{E}^{\mathbb{Q}^0}\left[e^{-r\tau}(1-R)V_{\tau}^+\right]&=(1-R)\mathbb{E}^{\mathbb{P}}\left[e^{-r\tau}V_{\tau}^+L^0_{\tau}\right]\\
&=(1-R)\int_0^{\infty}\mathbb{E}^{\mathbb{P}}\left[e^{-rt}V^+_tL^0_{t}\right]f^{\mathbb{P}}(t)dt\\
&=(1-R)\int_0^{\infty}\mathbb{E}^{\mathbb{Q}^*}\left[e^{-rt}V^+_t\right]f^{\mathbb{P}}(t)dt\\
&=(1-R)\int_0^{\infty}\mathbb{E}^{\mathbb{Q}^*}\left[e^{-rt}V^+_t\right]\lambda^{\mathbb{P}}e^{-\lambda^{\mathbb{P}}t} dt,
\end{split}
\]
where the second equality is by the change of equivalent martingale measure and the third equality is by the fact that $L_t^0$ is independent of $\tau$ and that $\tau$ is independent of the Brownian motion $W$. The equivalent martingale measure $\mathbb{Q}^*$ is given by the Radon-Nikodym derivative $Z^*$ defined in (\ref{z}).

We note that in the case of a counterparty with liquid CDS we would find the implied risk-neutral default intensity $\lambda^{\mathbb{Q}^H}$ from the CDS spreads and use it to calculate the CVA, i.e. we would have
\[CVA=\mathbb{E}^{\mathbb{Q}^H}\left[e^{-r\tau}(1-R)V_{\tau}^+\right]=(1-R)\int_0^{\infty}\mathbb{E}^{\mathbb{Q}^*}\left[e^{-rt}V^+_t\right]\lambda^{\mathbb{Q}^H}e^{-\lambda^{\mathbb{Q}^H}t} dt.\]

\subsection{Default probabilities based on Expected Default Frequencies}
\label{method2}
We choose a set $J$ of firms that have both: liquid 5-year CDSs and expected default frequencies (EDF) published on Moody's CreditEdge portal on a given time interval $I$ given in days. EDF is a firm specific forward-looking measure of real-world probability of default that is calculated by Moody's based on the Kealhofer-McQuown-Vasicek (KMV) model. The main idea is that firm’s equity can be seen as a call option on the underlying asset with the strike price equal to the face value of the firm’s debt. Then a mapping is used to transfer the distance-to-default to historical defaults.

As it was discussed in \cite{brigo}, the CDS spreads are not pure measures of credit risk and a methodology is needed to disentangle the liquidity premium from them. Hence, the cleaned CDS spreads will be used to derive the implied probabilities of default, where a cleaned CDS spread means a CDS spread without the liquidity premium. Then, for every $i\in I$ for every firm $j\in J$ we bootstrap the default intensity from the cleaned CDS spread and denote it by $\lambda^{CDS}_{ij}.$ Moreover, we can calculate the EDF-implied default intensity $\lambda_{ij}^{EDF}$ by using the following formula
\[\lambda_{ij}^{EDF} = -\ln(1-p_{if}^{EDF}),\]
where $p_{if}^{EDF}$ is the $EDF$-implied 1-year default probability.

Hence, we have a sequence of pairs $(\lambda^{EDF}_{ij},\lambda^{CDS}_{ij})_{(i\in I, j\in J)}.$

Similarly to \cite{c4} we assume a simple linear model between the natural logarithm of the EDF-implied default intensity $\lambda_{ij}^{EDF}$ and the natural logarithm of the CDS-implied default intensities $\lambda^{CDS}_{ij}$, i.e.
\[\ln(\lambda^{EDF}_{ij})=\gamma_i^0+\gamma_i^1\ln(\lambda^{CDS}_{ij})+\epsilon_{i},\]
where $\epsilon_{i}$ is a standard normal random variable.

Then, we use the obtained parameters $\gamma^{0}_i$ and $\gamma^{1}_i$ to calculate the default intensity for a counterparty that does not have a liquid CDS in the following way: Let $C$ denote the set of Swedbank's counterparties without liquid CDS. Firstly, we calculate a cleaned CDS proxy as discussed in Appendix \ref{nomura} and then we assume that the above mentioned linear relationship  holds also for the pairs $(\lambda^{CDS^{proxy}_{ic}},\lambda_{ic}^{\mathbb{P}})_{i\in I, c\in C},$ i.e. that we have
\[\ln(\lambda_{ic}^{\mathbb{P}})=\gamma_i^{0}+\gamma_i^{1}\ln(\lambda^{CDS^{proxy}}_{ic})+\epsilon_i,\]
where for every $i\in I$ and $c\in C$ we have that $\lambda_{ic}^{\mathbb{P}}$ is the real-world intensity.

Hence, for every day $i\in I$ and for every counterparty $c\in C,$ the cumulative distribution function $p_{ic}(t)$ of the default time $\tau$ is calculated by the following formula
\[p_{ic}(t)=\mathbb{P}(\tau\leq t)=1-e^{-\lambda_{ic}^{\mathbb{P}} t},\]
where
\[\lambda_{ic}^{\mathbb{P}} =e^{\gamma^{0}_i+\gamma^{1}_i\ln(\lambda^{CDS^{proxy}}_{ic})}.\]

\section{Appendix: The Nomura method}
\label{nomura}
The cross-sectional methodology for calculating proxy CDS spreads is based on \cite{c5} and can be summarized as follows. We fix a set of ratings, regions and sectors and choose a universe of counterparties with liquid CDS spreads. We label every counterparty with a rating, region and sector. The proxy CDS spread for a given counterparty is
\[S_i^{proxy}=M_{global}M_{rating(i)}M_{region(i)}M_{sector(i)},\]
where:
\begin{itemize}
\item $M_{global}$ is a global factor
\item $M_{rating(i)}$ is the factor for the rating of counterparty $i,$
\item $M_{region(i)}$ is the factor for the region of counterparty $i,$
\item $M_{sector(i)}$ is the factor for the sector of counterparty $i,$
\end{itemize}
For example, for a counterparty from North America with sector $FIN$ and rating $AA,$ we would have
\[S_i^{proxy}=M_{global}M_{AA}M_{North America}M_{FIN}.\]
As it was discussed in \cite{brigo}, the CDS spreads are not pure measures of credit risk and a methodology is needed to disentangle the liquidity premium from them. Hence, the cleaned CDS proxy spreads should be used to derive the implied probabilities of default, where a cleaned CDS proxy spread means a CDS proxy spread without the liquidity premium.
\subsubsection{A method for finding the factors of the cross-sectional method}
To find the factors of the cross-sectional method we can follow the following steps.
\begin{itemize}
\item Enumerate all the factors starting with $M_{global},$
\item Let $n$ be the total number of all factors (for example for $7$ regions, $7$ ratings and $11$ sectors we have $n = 26$)
\item Denote $y_i = \ln(S_i^{proxy})$ and $x_j = \ln(M_j)$
\item Let $A=[A_{ij}]$ be a matrix of 0s and 1s
\end{itemize}
Then we can write the model as
\[y_i = \sum_{j=1}^n A_{ij} x_j.\]
We want to find the optimal $x$ that makes the proxy spreads as close as possible to the market spreads.
We can define "as close as possible" to mean "minimizing total squared difference in log spreads."
Then finding the optimal $x$ simply means performing a linear regression.

\section{Estimating real-world default probabilities from the CDS market}
\label{default_prob}

The idea can be summarized in the following steps.
\begin{itemize}
\item[Step 1] Fix a set of ratings, regions and sectors
\item[Step 2] Choose a universe of firms with liquid CDS
\item[Step 3] Label every firm with a rating, region and sector chosen in Step 1
\item[Step 4] Label every counterparty with a rating, region and sector chosen in Step 1
\item[Step 5] Calculate a CDS proxy for every counterparty with a cross-sectional method (see Appendix \ref{nomura})
\item[Step 6] Bootstrap default intensity from the CDS proxy for every counterparty
\item[Step 7] Calculate the real-world default intensity $\lambda^{\mathbb{P}}$
\item[Step 8] Use the default intensity $\lambda^{\mathbb{P}}$ to calculate default probabilities
\end{itemize}
In Appendix \ref{nomura} we present a method for calculating CDS proxy spreads what are needed in the Step 5 above and in Subsection \ref{method2} we present a method that can be used for calculating real-world default probabilities based on the CDS proxy spreads and Expected Default Frequencies (EDF).

As an example we present in the following section a methodology that can be used for estimating real-world default probabilities from the CDS market with the use of Expected Default Frequencies (EDF).

\end{document}